\documentclass{DISproc-n}

\begin{document}
\title{Inclusive $D$-Meson Production at the 
LHC\footnote{
  Talk presented at XX.\ International Workshop on Deep-Inelastic 
  Scattering and Related Subjects, 26-30 March 2012, University 
  of Bonn. Work in collaboration with B.\ A.\ Kniehl, G.\ Kramer, 
  and I.\ Schienbein.
  }
}

\author{{\slshape Hubert Spiesberger}\\[1ex]
Johannes-Gutenberg-Universit\"at, 55099 Mainz, Germany}



\maketitle

\begin{abstract}
  I present predictions for the inclusive production of $D$ mesons at 
  the CERN LHC in the general-mass variable-flavor-number scheme at 
  next-to-leading order. Numerical results are compared to data where 
  available, and uncertainties due to scale variations, parton 
  distribution functions and charm mass are discussed. I point out 
  that measurements at large rapidity have the potential to pin down 
  models of intrinsic charm.
\end{abstract}

$D$-meson production at the LHC was studied by the ALICE 
\cite{ALICE:2011aa}, ATLAS \cite{atlas-note-2011}, and LHCb 
Collaborations \cite{lhcb-confnote-2011}. Here I present 
predictions for the inclusive production of $D$ mesons
at the LHC within the general-mass variable-flavor-number 
scheme (GM-VFNS) \cite{KKSS}. More results and additional 
details of the calculation can be found in \cite{KKSS-d4lhc}. 
In a recent paper \cite{Kniehl:2011bk}, we have also considered 
the inclusive production of $B$ mesons, for which experimental 
results from the CMS Collaboration are available \cite{CMS}.
For an alternative approach, see Ref.~\cite{FONLL-12}.

\begin{wrapfigure}{r}{0.45\textwidth}
  \centering
  \includegraphics[width=0.4\textwidth]{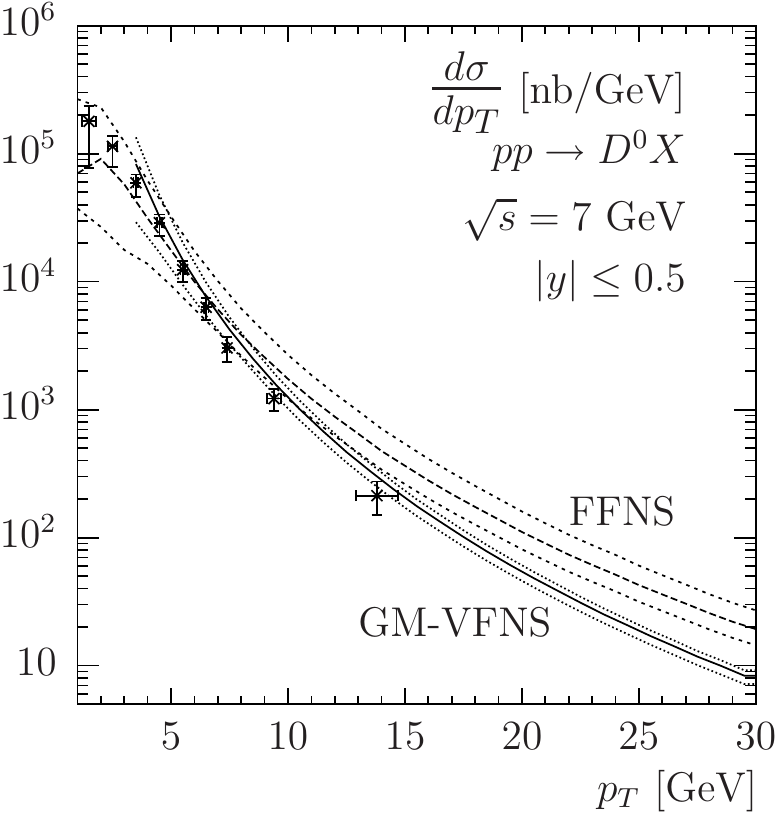}
  \caption{
  $\mathrm{d}\sigma/\mathrm{d}p_T$ for $p+p \to D^0+X$ integrated over 
  rapidity in the range $-0.5 \leq y \leq 0.5$ for $\sqrt{s}=7$~TeV 
  at NLO in the GM-VFNS (solid line) and the FFNS (dashed line). 
  Dotted lines describe the corresponding error bands from scale 
  variations. The ALICE data were taken from \cite{ALICE:2011aa}.
  }
  \label{Fig:HS-1}
\end{wrapfigure}
Figure \ref{Fig:HS-1} shows a comparison of the GM-VFNS predictions 
for the transverse momentum distribution with data from ALICE. 
Here the renormalization ($\mu_R$) and factorization scales for 
initial state ($\mu_I$) and final state ($\mu_F$) singularities 
are fixed by $\mu_i = \xi_i \sqrt{p_T^2+m_c^2}$, where $m_c$ is the 
charm quark mass, and the scale parameters $\xi_i$ ($i = R, F, I$) 
are varied about the default values of 1 by factors of 2 up and down 
to obtain an estimate of a theory uncertainty band (dotted lines in 
the figure). The data are reasonably well described by theory at the 
larger values of $p_T$, where data are available, but due to the 
choice $\xi_i = 1$, theory starts to overshoot at $p_T < 5$ GeV. 
There, the fixed flavor number scheme (FFNS) \cite{ffns-theory} works 
better (see the dashed lines in Fig.~\ref{Fig:HS-1}). The GM-VFNS is 
preferred at large $p_T$ since it includes resummed contributions from 
large logarithms by virtue of the DGLAP evolution equations for the 
parton distribution (PDFs) and fragmentation functions (FFs). The 
GM-VFNS also predicts smaller scale uncertainties than the FFNS. We 
have used CTEQ6.6 PDFs \cite{CTEQ6.6} and, in the case of the GM-VFNS, 
FFs of Ref.~\cite{Kneesch:2007ey}. The FFNS calculation is performed 
without including a FF; the transition from the charm quark to the 
charmed meson is taken into account by multiplying the parton level 
result with the branching ratio $BR(c\rightarrow D^0) = 0.628$.

\begin{figure}[t]
  \centering
  \includegraphics[width=0.32\textwidth]{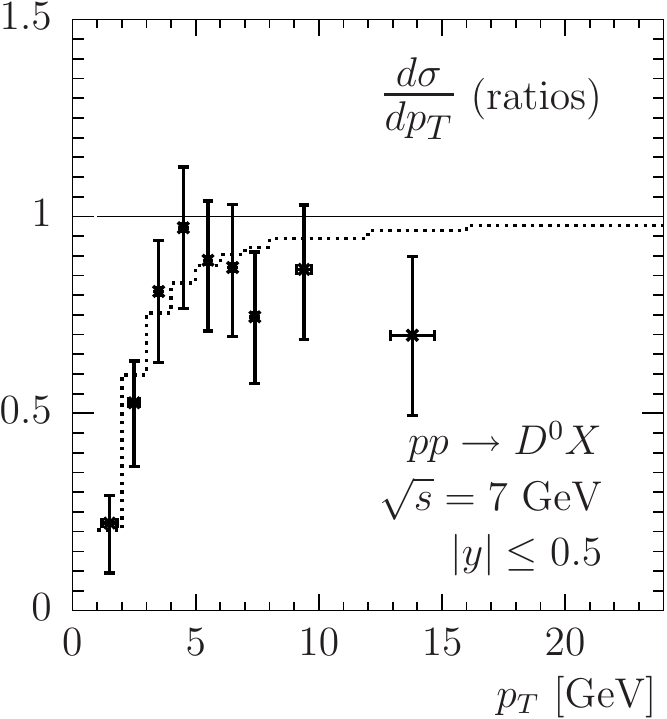}
  \includegraphics[width=0.32\textwidth]{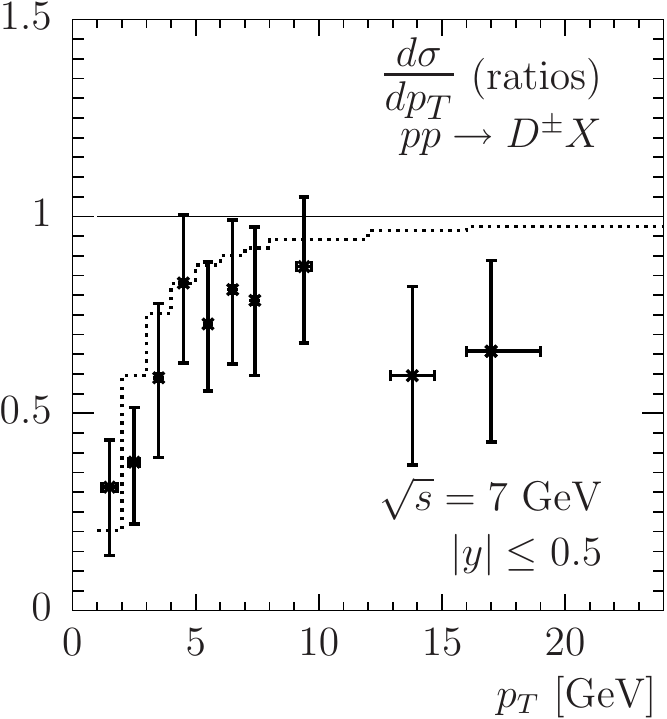}
  \includegraphics[width=0.32\textwidth]{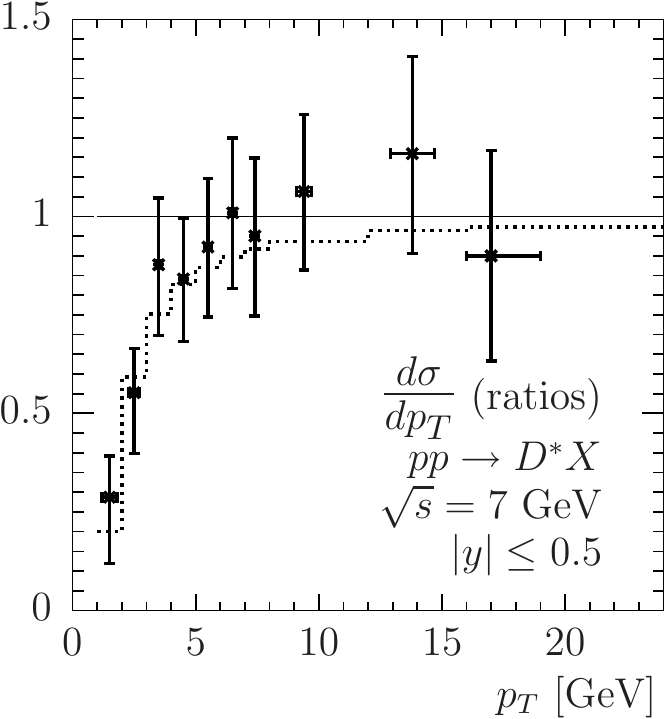}
  \caption{
  Ratios of $d\sigma/dp_T$ for $D$-meson production at ALICE at 
  $\sqrt{s}=7$~TeV using $\xi_I = \xi_F = 0.8$ and $\xi_R = 1$. All 
  cross sections and the data from Ref.\ \cite{ALICE:2011aa} are 
  normalized to the GM-VFNS prediction with $\xi_i=1$. The PDFs are 
  taken from MSTW08-NLO \cite{Martin:2009iq} and the charm quark 
  mass is $m_c = 1.5$ GeV.
  }
  \label{Fig:HS-2}
\end{figure}

\begin{figure}[b!]
  \centering
  \includegraphics[width=0.32\textwidth]{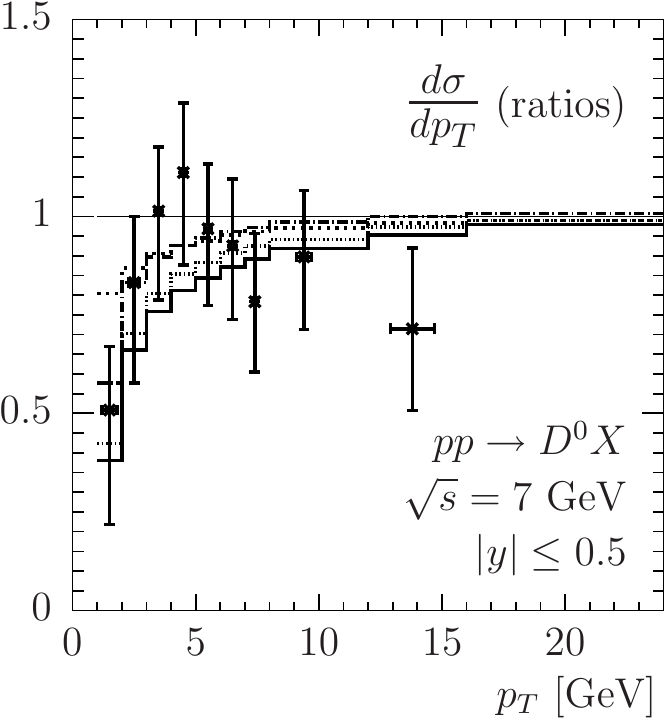}
  \includegraphics[width=0.32\textwidth]{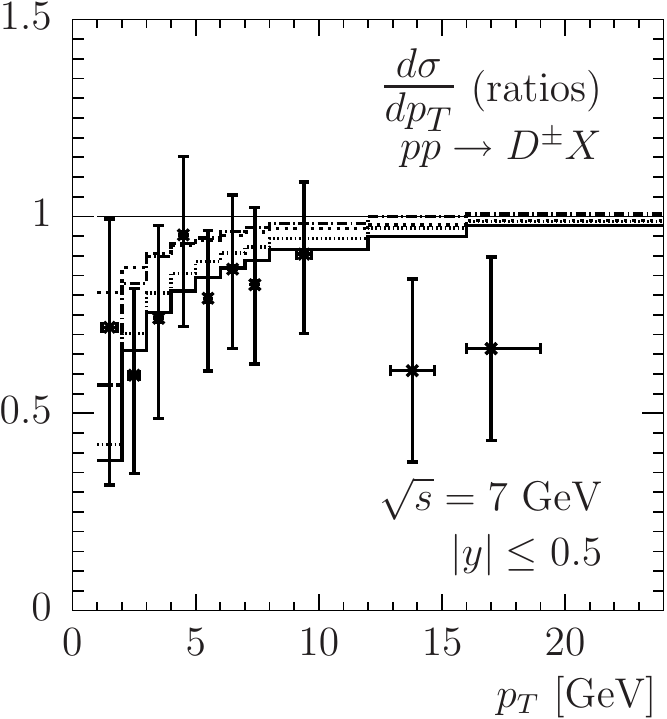}
  \includegraphics[width=0.32\textwidth]{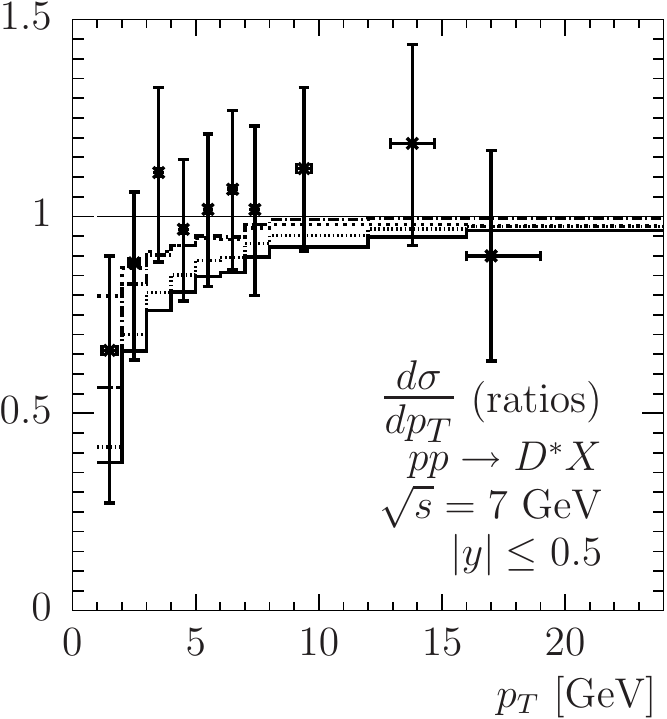}
  \caption{
  $d\sigma/dp_T$ for $D$-meson production at ALICE at $\sqrt{s}=7$~TeV 
  for different PDFs. All cross sections are calculated with 
  $\xi_I = \xi_F = 0.7$, $\xi_R = 1$ and normalized to the GM-VFNS 
  prediction with CTEQ6.6 PDFs. The histograms from top down 
  correspond to CT10 \cite{Lai:2010vv}, HERAPDF 1.5 (NLO) 
  \cite{herapdf15}, MSTW08-NLO \cite{Martin:2009iq} and NNPDF 2.1 
  \cite{Ball:2011mu}. 
  }
  \label{Fig:HS-3}
\end{figure}

The uncertainties due to variations of the factorization scales 
are dominant. It is interesting to see that the scale parameters 
can be chosen to bring the GM-VFNS predictions into agreement with 
the data also at low values of $p_T$. This is shown in 
Fig.~\ref{Fig:HS-2} for MSTW08-NLO PDFs \cite{Martin:2009iq} and 
using $m_c = 1.5$ GeV for the charm quark mass. The differential 
cross sections $d\sigma/dp_T$ are shown here for $\xi_I = \xi_F = 
0.8$, $\xi_R = 1$ in $p_T$ bins and compared with data points from 
the ALICE collaboration \cite{ALICE:2011aa}. All results are 
normalized to the GM-VFNS prediction with $\xi_i = 1$. One can see 
that a proper choice of the factorization scales can help to ensure 
that the resummed contributions due to incoming heavy quarks, and 
those due to gluon fragmentation, fade out in a controlled manner 
as $p_T/m \rightarrow 0$, i.e.\ in the kinematic region where the 
FFNS should be appropriate.

\begin{wrapfigure}{r}{0.4\textwidth}
  \centering
  \includegraphics[width=0.32\textwidth]{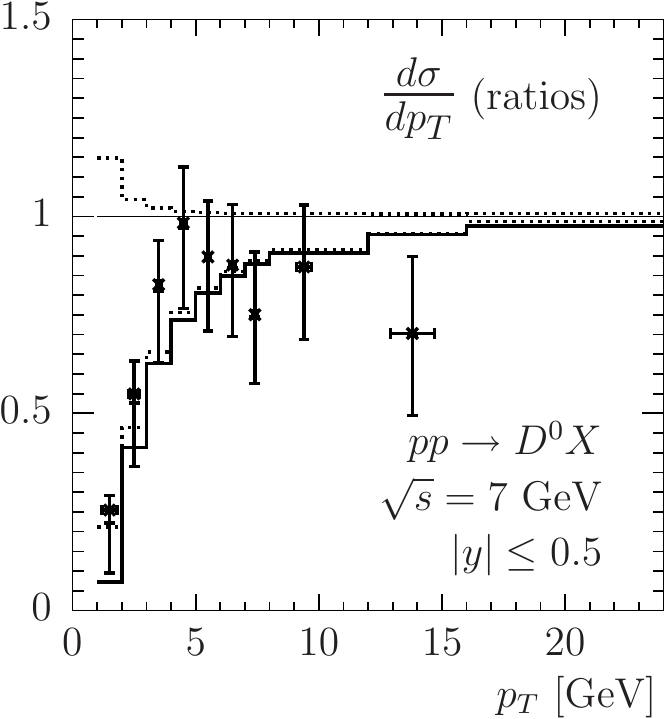}
  \caption{
  $d\sigma/dp_T$ for $D^0$-meson production at ALICE \cite{ALICE:2011aa} 
  with MSTW08-NLO PDFs \cite{Martin:2009iq} normalized to the GM-VFNS 
  prediction with $\xi_{I,F,R} = 1$ and $m_c = 1.5$ GeV. The 
  lower two histograms are obtained using $\xi_{I,F} = 0.7$ and 
  $\xi_R = 1$. For the two dashed histograms, $m_c = 1.4$ GeV was 
  used in both the hard scattering matrix elements and in the 
  MSTW08-NLO PDFs.
  }
  \label{Fig:HS-4}
\end{wrapfigure}
In Figure \ref{Fig:HS-3} an attempt is made to show the uncertainties 
coming from using different PDF input. The results for most of the 
bins lie within the error bars of the experimental data and do not 
prefer one PDF set over another. Actually, due to different values of 
$m_c$ used in the PDF fits and the corresponding different lengths in 
the evolution path from the charm production threshold up to $\mu_I$, 
there is some residual $m_c$ dependence of the predicted cross sections 
at low values of $p_T$. 
The value $m_c = 1.5$ GeV used in our calculation agrees with the one 
in the fragmentation functions of Ref.~\cite{Kneesch:2007ey}, but not 
with the one in the parton distribution functions used here. While the 
CTEQ6.6 and CT10 sets use $m_c = 1.3$ GeV, in the MSTW08-NLO, NNPDF 2.1, 
and HERAPDF 1.5 (NLO) sets $m_c = 1.4$ GeV was chosen. A consistent 
calculation would require the same value of $m_c$ in all components 
of the cross section formula. However, separate fits of the fragmentation 
functions for different values of $m_c$ are not available. The dependence 
on the heavy quark mass is, however, not very strong and non-negligible 
only in the low $p_T$ range, see Fig.~\ref{Fig:HS-4}.

\begin{figure}[t!]
  \centering
  \parbox{0.35\textwidth}{
  \includegraphics[width=0.35\textwidth]{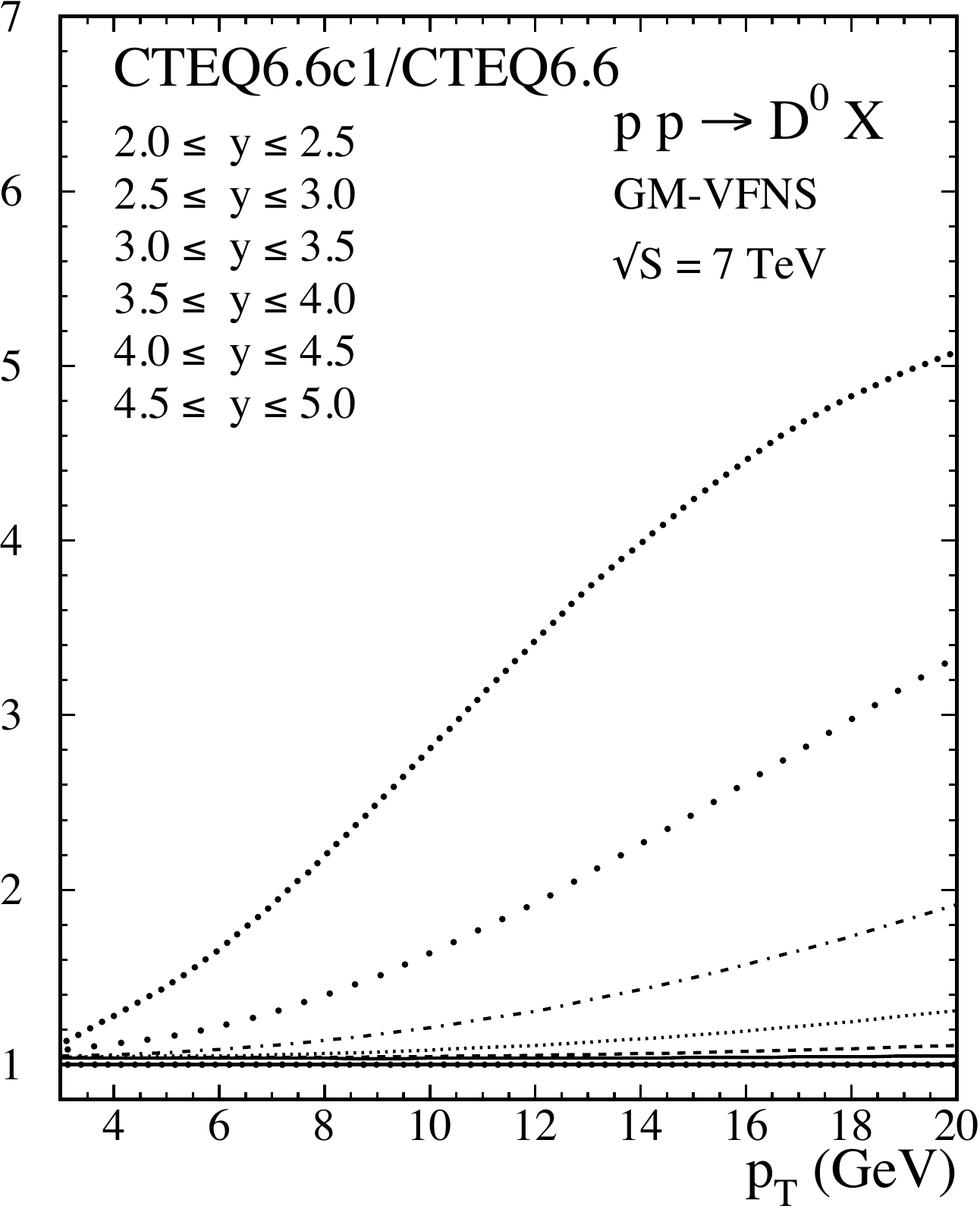}}
  \hspace*{6mm}
  \raisebox{-1mm}{\parbox{0.36\textwidth}{
  \includegraphics[width=0.362\textwidth]{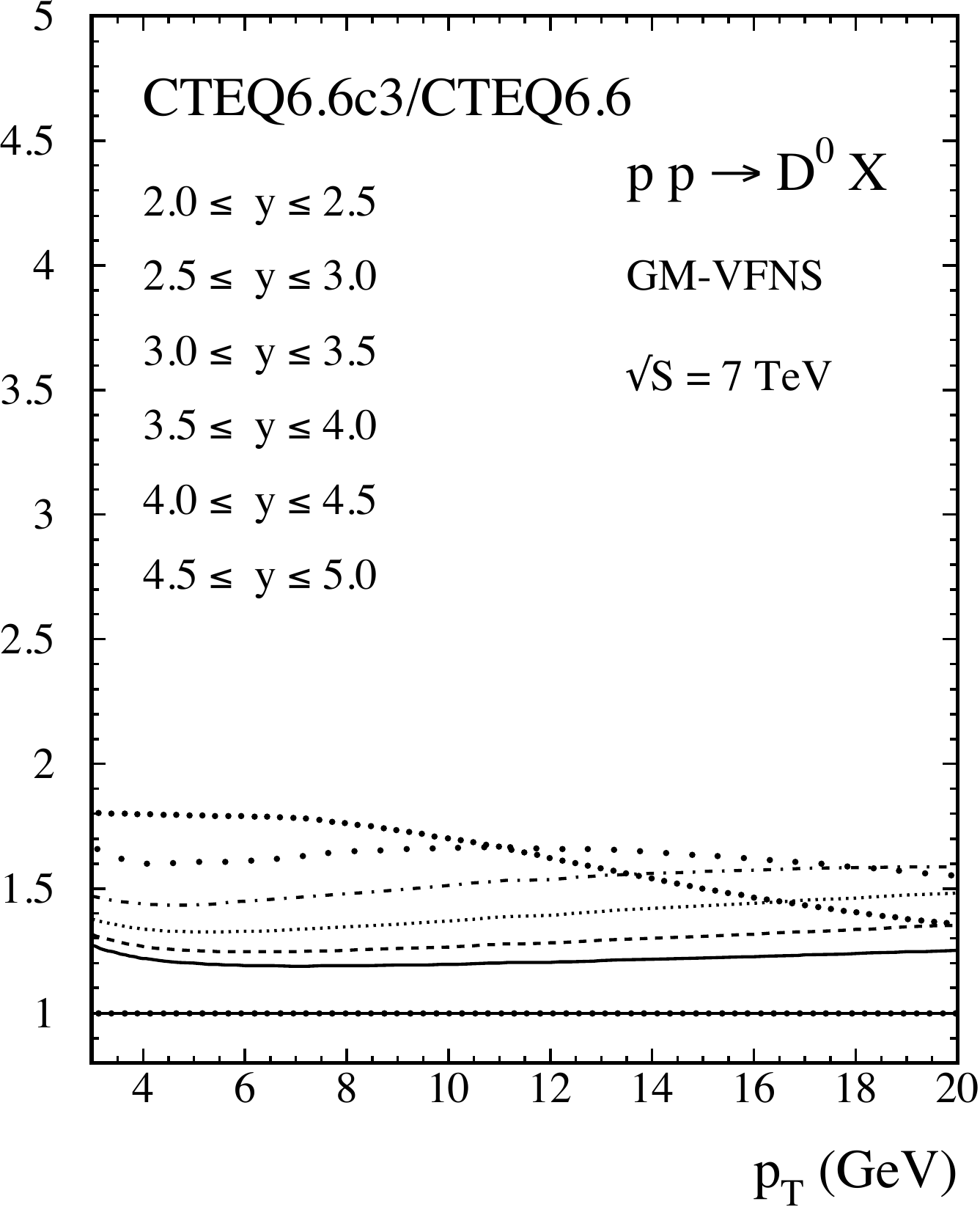}}}
  \caption{
  Ratio of the $p_T$ distributions $\mathrm{d}\sigma/\mathrm{d}p_T$ 
  for $p+p \to D^0 + X$ at NLO in the GM-VFNS at $\sqrt{s} = 7$ TeV, 
  using two different models of intrinsic charm: 
  (a) BHPS model with 3.5\,\% $(c+\overline{c})$ content (at $\mu_F 
  = 1.3$ GeV), 
  (b) model with a high strength sea-like charm component. 
  The FFs are taken from Ref.\ \cite{Kneesch:2007ey}. The various 
  lines represent the default predictions for $\xi_R = \xi_I = \xi_F 
  = 1$, integrated over the rapidity regions indicated in the figures 
  (larger rapidities correspond to larger cross section ratios 
  everywhere in (a) and at small $p_T$ in (b)).
  }
  \label{Fig:HS-5}
\end{figure}

Non-perturbative contributions to the charm quark content of the 
proton may lead to enhanced charm parton distributions $c(x,\mu_F)$ 
at $x > 0.1$. This can become visible in the cross section for $D$ 
meson production at large rapidities. Parametrizations of this 
so-called intrinsic charm are available from the CTEQ collaboration, 
based on various models and compatible with the global data samples. 
In Ref.~\cite{Kniehl:2009ar}, we have studied the impact of these 
models on possible measurements at the Tevatron and at BNL RHIC. 
Here, I present results of a calculation using the parametrization 
CTEQ6.6 \cite{CTEQ6.6} to obtain an estimate of the expected relative 
enhancements of the $p_T$ distributions in bins of rapidity. 
Figure~\ref{Fig:HS-5} shows typical results for $D^0$ production; 
for other $D$ mesons, the results are very similar. Two models 
have been selected among the possible options in CTEQ6.6 (see Ref.\ 
\cite{Pumplin:2007wg} for details): Fig.\ \ref{Fig:HS-5}a shows the 
calculation using the BHPS model with a 3.5\,\% $(c + \overline{c})$ 
content in the proton (at the scale $\mu_F = 1.3$ GeV), Fig.\ 
\ref{Fig:HS-5}b refers to the model of a high strength sea-like 
charm component. In both cases, one observes large enhancements, 
increasing with rapidity, and in the first model also with $p_T$. 
Thus one can expect that forthcoming data from the LHCb experiment 
should be able to exclude or narrow down models for intrinsic charm. 

{\raggedright
\begin{footnotesize}


\end{footnotesize}
}


\end{document}